\documentclass[usenatbib,useAMS]{mn2e}  
\usepackage{times}
\usepackage{epsf}
\usepackage{amssymb}

\title[The chemical enrichment of elliptical galaxies]{
The metal enrichment of 
elliptical galaxies in hierarchical galaxy formation models 
}

\author[M. Nagashima et al.]{Masahiro Nagashima
\thanks{E-mail: masa@scphys.kyoto-u.ac.jp (MN).},$^{1,2}$
Cedric G. Lacey,$^{2}$
Takashi Okamoto,$^{2,3}$
Carlton M. Baugh,$^{2}$
\newauthor{Carlos S. Frenk,$^{2}$ and Shaun Cole$^{2}$}\\
$^{1}$Department of Physics, Graduate School of Science,
Kyoto University, Sakyo-ku, Kyoto 606-8502, Japan\\
$^{2}$Department of Physics, University of Durham, South Road, Durham DH1 3LE,
United Kingdom\\
$^{3}$National Astronomical Observatory, National Institute of Natural
Science, Mitaka, Tokyo 181-8588, Japan
}

\begin{document}   
\maketitle   

\begin{abstract}   
We investigate the metal enrichment of elliptical galaxies in the
framework of hierarchical models of galaxy formation. The
semi-analytical model we use in this {\it Letter}, which has been used
to study the metal enrichment of the intracluster medium (ICM) by
Nagashima et al., includes the effects of flows of gas and metals both
into and out of galaxies and the processes of metal enrichment due to
both type Ia and type II supernovae. We adopt a solar neighbourhood IMF
for star formation in discs, but consider models in which starbursts
have either a solar neighbourhood IMF or a top-heavy IMF. We find that
the $\alpha$-element abundance in ellipticals is consistent with
observed values only if the top-heavy IMF is used.  This result is
consistent with our previous study on the metal enrichment of the ICM.
We also discuss the abundance ratio of $\alpha$ elements to iron as a
function of velocity dispersion and metallicity. We find that models
with a top-heavy IMF match the $\alpha$/Fe ratios observed in typical
$L_*$ ellipticals, but none of the models reproduce the observed
increase of $\alpha$/Fe with velocity dispersion.

\end{abstract}   
   

\section{INTRODUCTION}   

The chemical abundances of elliptical galaxies have long been known
to provide important constraints on theoretical models of their
formation. The metallicities of the stellar populations in elliptical
galaxies are estimated observationally from the strength of stellar
absorption features in their integrated spectra. Such measurements
indicated an increase of the overall metal abundance with galaxy
luminosity and velocity dispersion \citep[e.g.][]{faber73, bender93},
and also an increase in the average Mg/Fe abundance ratio
\citep[e.g.][]{oconnell76, worthey92, jorgensen99}, with the central
regions of most bright ellipticals having above-solar total
metallicities and Mg/Fe ratios (the so-called {\em
$\alpha$-enhancement}). The detailed interpretation of line strengths
in terms of chemical abundances is, however, complicated, due to the
non-solar abundance ratios, and to the dependence of absorption line
strengths on both metallicity and age. Detailed stellar population
synthesis models have been developed to relate measured absorption
line strengths to ages and metallicities; older models
\citep[e.g.][]{worthey94} assumed solar abundance ratios, but newer
models \citep[e.g.][]{trager00b, thomas03} allow for non-solar
ratios. Nonetheless, there remain significant uncertainties in going
from measured line strengths to abundances of different elements.

Most theoretical modelling of chemical abundances in ellipticals has
been in the framework of the {\em monolithic collapse} or {\em single
burst} model, in which all of the stars form at high redshift in a
single burst of short duration, usually terminated by ejection of the
remaining gas by a wind, following which the stars evolve
passively \citep[e.g.][]{larson75, ay87}. Subsequent works
\citep[e.g.][]{matteucci94, tgb99} have shown that it is possible
to explain the trends of both increasing total metallicity and 
increasing Mg/Fe ratio with galaxy mass within this framework,
provided that the star formation timescale of the initial burst (a free
parameter) is chosen to have a suitable dependence on galaxy mass.

However, there is now overwhelming evidence that structure in the
universe formed by hierarchical clustering of a dominant Cold Dark
Matter (CDM) component, and strong evidence that galaxy mergers played
an important role in the formation of elliptical galaxies. Galaxy
formation within this {\em hierarchical merger} framework has been
studied extensively using semi-analytical models \citep[e.g.][]{wf91, kwg93,
cafnz94, clbf00, sp99, ntgy01, mcfgp02, hatton03}. There have, however,
been very few detailed studies of the chemical evolution of elliptical
galaxies within this framework. \citet{clbf00} and \citet{ny04}
computed the relation between total metallicity and luminosity for
ellipticals, while \citet{kc98} calculated the relation between Mg
line strength and galaxy velocity dispersion, but these calculations
all used the instantaneous recycling approximation for chemical
evolution, with fixed solar abundance ratios built in. By
construction, such models are unable to explain variations in
abundance ratios such as Mg/Fe. \citet{t99} and \citet{tk99}
calculated the separate evolution of [Fe/H] and [Mg/Fe], including
both the prompt metal ejection by Type~II supernovae (SNe~II) and the
long-delayed ejection by Type~Ia supernovae (SNe~Ia), based on star
formation histories taken from a semi-analytical model. They found in
their model that the average [Mg/Fe] actually decreased with
increasing luminosity, in apparent conflict with observational
data. However, they calculated chemical evolution using a {\em
closed-box} model, which ignored the effects of galaxy mergers and the
transport of metals into and out of galaxies by gas inflows and
outflows, all of which form part of standard semi-analytical models.

In this paper we present the first calculations of chemical abundances
in elliptical galaxies in a semi-analytical model which includes
enrichment by both SNe~Ia and SNe~II within a fully consistent
framework of hierarchical galaxy formation, i.e. including gas inflows
and outflows and galaxy mergers. This enables us to investigate
whether hierarchical merger models are able to reproduce the observed
variations in the ratio, $\alpha$/Fe, of $\alpha$-elements to iron. We
base our calculations on the semi-analytical model described in
\citet{baugh05}, which assumes that stars formed in bursts triggered
by galaxy mergers have a top-heavy IMF, while stars formed quiescently
in discs have a \citet{k83} IMF, similar to the solar
neighbourhood. The top-heavy IMF in bursts was introduced in order
that the model fit the number counts of faint sub-mm
galaxies. \citet{nlbfc05} showed that this same model explained the
observed abundances in the intracluster medium (ICM) in galaxy
clusters, when enrichment by SNe~Ia was included. In the present
paper, we compare predictions from the same model with observed
abundances of elliptical galaxies. \citet{no04} have already shown
that a similar hierarchical model is able to explain the abundance
patterns in disc galaxies like the Milky Way.

The general idea that some or all of the stars in elliptical galaxies
may have formed with an IMF which was top-heavy relative to that in
the solar neighbourhood is not a new one. \citet{matteucci94},
\citet{gibson97} and \citet{tgb99} all proposed a top-heavy IMF as one
possible explanation for the abundance patterns observed in
ellipticals. The model presented here is novel in that it is
based within a fully consistent hierarchical framework, and in that
the parameters of the top-heavy IMF and starbursts have been chosen in
advance, independently of any metallicity constraints. We note that
there is some more direct evidence for a top-heavy IMF in bursts from
observations of the starburst galaxy M82 \citep{takp97,sg01}. 

Thus, the purpose of this {\it Letter} is to investigate theoretical
predictions for the metallicities of elliptical galaxies from the same
hierarchical galaxy formation model as we used to study the enrichment
of the ICM in \citet{nlbfc05}, and to make a preliminary comparison
with observational data. In the next section we briefly summarise our
model.  We present the results from our model in Section 3 and our
conclusions in Section 4.  In the following, the cosmological
parameters are fixed to be $\Omega_{0}=0.3, \Omega_{\Lambda}=0.7,
h\equiv H_{0}/100$ km~s$^{-1}$Mpc$^{-1}$ = 0.7 and $\sigma_{8}=0.93$.

\section{MODEL}
We adopt the same model as used in \citet{nlbfc05}, which is an
extension of the {\sc galform} semi-analytical galaxy formation model
\citep{clbf00, baugh05} to include both prompt metal enrichment by
SNe~II and delayed enrichment by SNe~Ia. The model, which will be
described in full in Lacey et al. (2005, in preparation), includes the
following physical processes: merging of dark haloes, radiative gas
cooling, star formation, SN feedback, mergers of galaxies due to
dynamical friction, starbursts triggered by such mergers, and stellar
population synthesis. It has been shown in our previous papers that this
model reproduces both the observed sub-mm source counts, and the
observed metal abundances and baryon fractions in the ICM, provided that
a top-heavy IMF is adopted for starbursts. We assume a \citet{k83} IMF
for quiescent star formation in galaxy discs, which has slope $x=0.4$
for $m<1 M_{\odot}$ and $x=1.5$ for $m>1 M_{\odot}$. The top-heavy IMF
has slope $x=0$ between 0.15 and 120$M_{\odot}$ (where $x=1.35$ for the
Salpeter IMF).

As shown by \citet{bbflbc03}, the original {\sc galform} model
\citep{clbf00}, when used with currently estimated cosmlogical
parameters, predicted an excessive number of very luminous galaxies at
the present epoch. \citet{nlbfc05} considered two alternative
solutions to this problem: (a) a {\em superwind} model [similar to
\citet{bbflbc03}], in which SN feedback expels gas from haloes, the
expelled gas later being recaptured once the halo grows up to a
circular velocity $V_{c}=V_{\rm recap}\equiv 600$ km~s$^{-1}$; (b) a
{\em conduction} model, in which the conductive heat flux from hot gas
in the outer parts of massive haloes prevents gas from cooling in the
centres of these haloes. Both models included a top-heavy IMF in
bursts, and were consistent with metallicities in the ICM of
clusters. We present results for both models in this paper. For
comparison purposes, we also show results for a {\em
superwind/Kennicutt} model, which is identical to our standard
superwind model except that a Kennicutt IMF is used throughout.

We calculate the evolution of the abundances of different elements in
the stars and gas using exactly the same prescription for chemical
evolution as in \citet{nlbfc05}. The rates and chemical yields of
SNe~Ia and SNe~II and gas restitution rates are computed consistently
with the adopted IMFs based on \citet{pcb98} and \citet{gr83}. We
compute a total metallicity [Z/H] and iron abundance [Fe/H] (both
being expressed as logarithmic values relative to solar). For
comparing with the observational data, we also compute an
$\alpha$-element abundance [$\alpha$/H] which includes O, Mg and other
elements which are assumed in observational analyses to vary in
step with these - see \citet{trager00a} and \citet{thomas03} for
details. The $\alpha$-abundance is dominated in mass fraction by
O. The $\alpha$-elements dominate the total metallicity, so [Z/H] is
close to [$\alpha$/H].

For comparing with observational data, we also need to compute stellar
velocity dispersions of elliptical galaxies in our model. We compute
the radii of stellar spheroids formed in mergers using the method
described in \citet{clbf00}, and we then compute the circular velocity
$V_{cS}$ at the half-mass radius of the spheroid including both its
self-gravity and the gravity of its dark halo. We then estimate the 1D
stellar velocity dispersion as $\sigma = V_{cS}/\sqrt{3}$.

\section{RESULTS}
\begin{figure*}
\epsfxsize=0.85\hsize
\hspace{-1cm}
\epsfbox{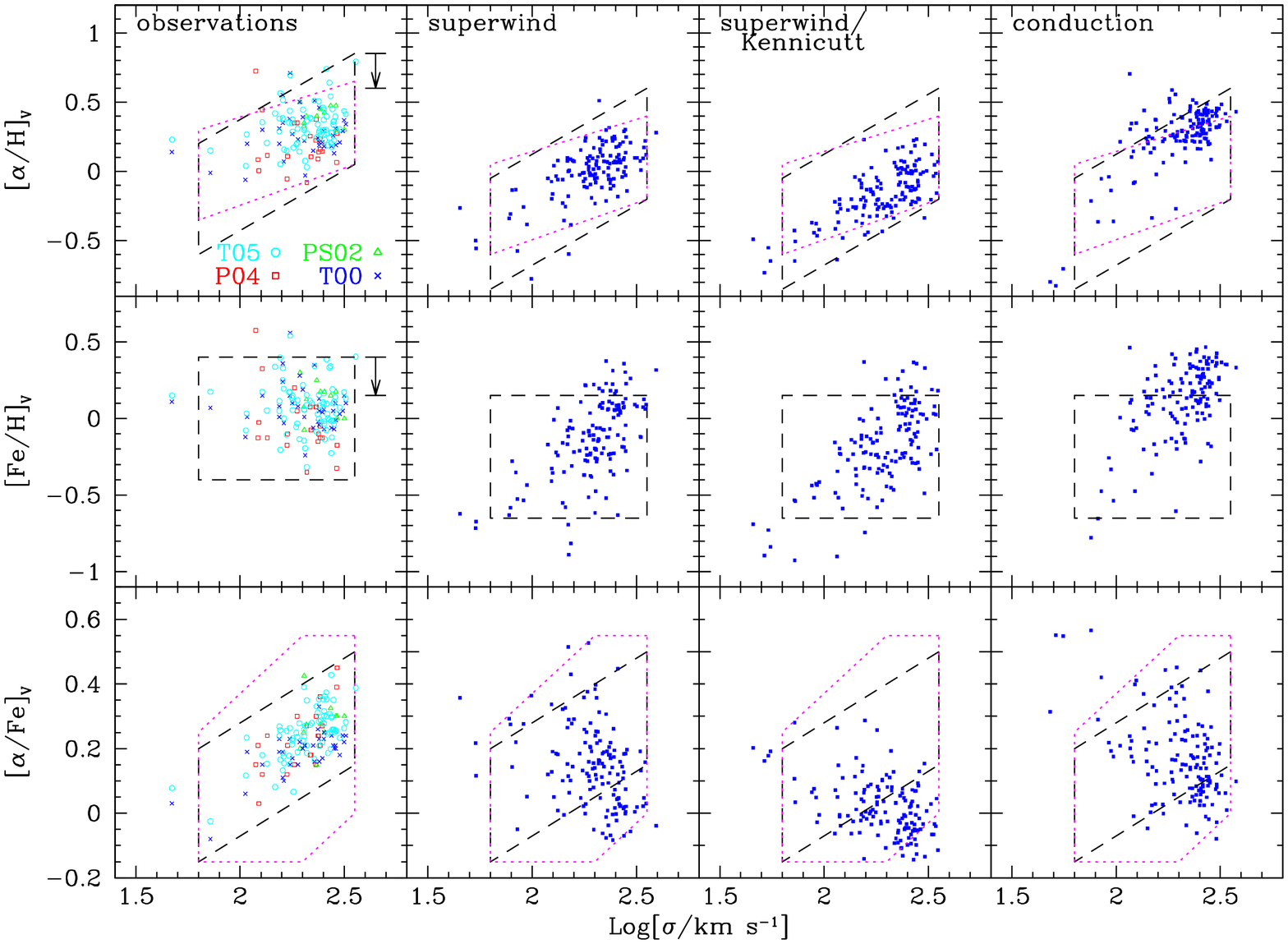}

\caption{Abundances of $\alpha$-elements, iron, and their ratio for
stars in elliptical galaxies, plotted against velocity dispersion
$\sigma$. Abundances are on a logarithmic scale, relative to solar. The
left-hand column shows observational data (not corrected for metallicity
gradients) as follows: \citet[][T05, {\it circles}]{tmbd05},
\citet[][P04, {\it squares}]{p04}, \citet[][PS02, {\it
triangles}]{ps02}, and \citet[][T00, {\it crosses}]{trager00a}. The
arrows show the estimated average corrections from central to global
values, which are -0.25 dex for [$\alpha$/H] and [Fe/H], and 0 for
[$\alpha$/Fe].  For reference, dashed boxes surrounding most of the
observational data are repeated across each row of panels.  In the right
3 columns, these boxes are shifted by the estimated central-to-global
corrections. The other 3 columns show predictions for the 3 different
models, as indicated. The model abundances are V-band
luminosity-weighted averages for each galaxy.  The dotted boxes indicate
how the observational data shift if the correction to [$\alpha$/Fe]
proposed by \citet{p04} is adopted.  }

\label{fig:fig1}
\end{figure*}

\begin{figure}
\epsfxsize=0.85\hsize
\epsfbox{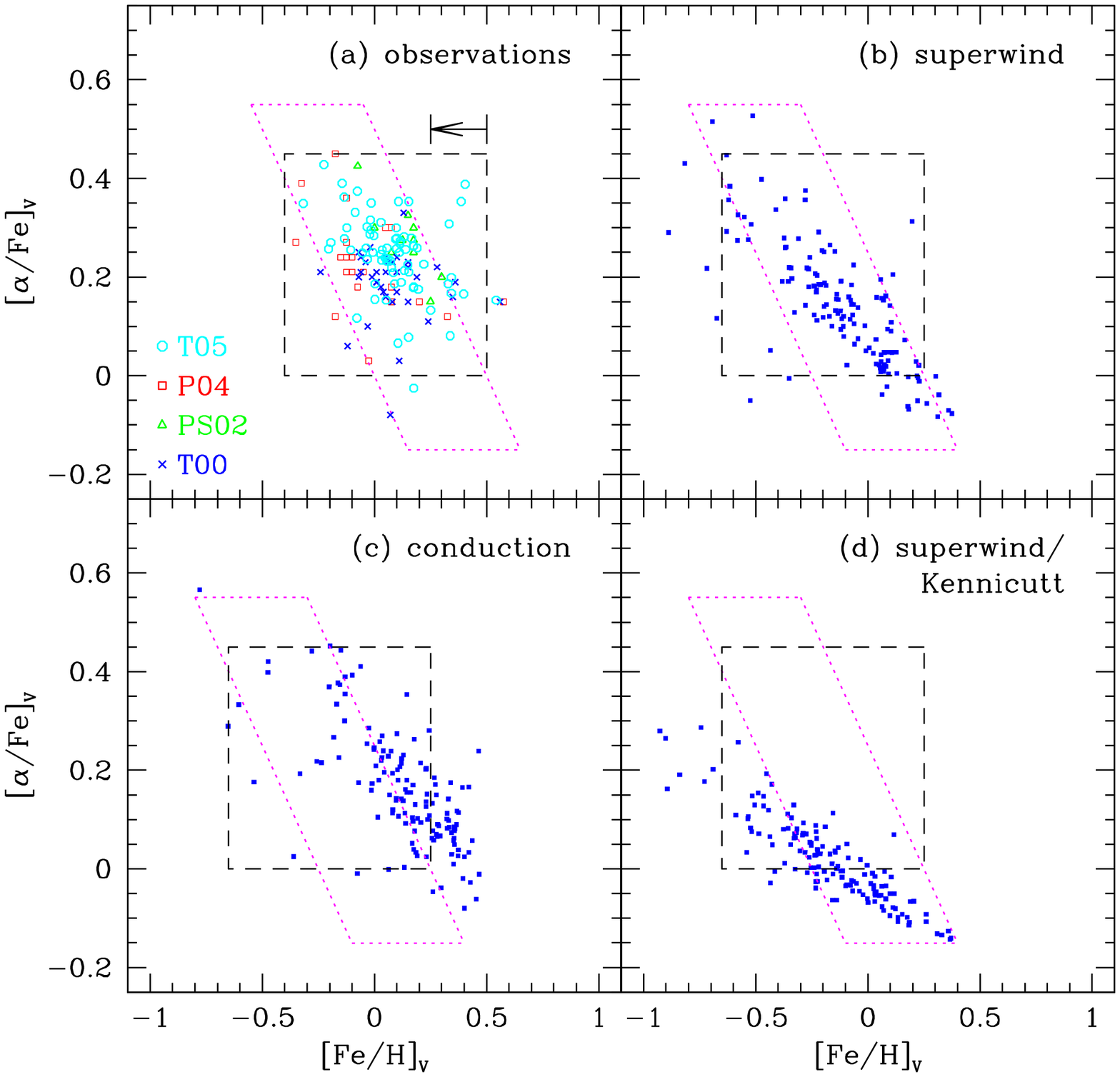}

\caption{The $\alpha$/Fe abundance ratio vs. Fe-abundance for stars in
elliptical galaxies. (a) Observational data, for the same galaxy samples
as in Fig.\ref{fig:fig2}.  Panels (b), (c) and (d) show predictions for
the three different models.  For reference, dashed boxes surrounding
most of the observational data are plotted in all panels. In the panels
showing the models, these boxes are shifted by the estimated
central-to-global correction for metallicity gradients.  The dotted
parallelogram indicates how the observational data shift if the
correction to [$\alpha$/Fe] proposed by \citet{p04} is adopted.}

\label{fig:fig2}
\end{figure}

Our predictions for the mean global stellar metallicities of
ellipticals for the three different models we consider are shown in
Figs.\ref{fig:fig1} and \ref{fig:fig2}. We select ellipticals in the
model as galaxies with B-band bulge-to-total light ratios
$(B/T)_B>0.6$. The metallicities plotted for model galaxies are
averages over all of the stars in the galaxy, weighted by the
contribution of stars of different ages and metallicities to the
V-band luminosity. This weighting is intended to mimic the measurement
of metallicities from spectral features around 5000\AA.

We have compiled a set of observational data to compare with the models,
by combining the samples of \citet{trager00a}, \citet{ps02}, \citet{p04}
and \citet{tmbd05}, and selecting ellipticals with measurement errors on
[$\alpha$/Fe] of less than 0.1 dex. This leaves us with about 140
ellipticals in a range of environments from groups to clusters, most of
them with velocity dispersions in the range $100 \la \sigma \la 300\
{\rm km~s^{-1}}$. (We note that a recent paper, \citet{nelan05}, has
reported very similar results to the above papers for much larger
samples of 4097 red-sequence galaxies in 93 low-redshift galaxy
clusters.)  In the above papers, metallicities have been derived from
measured absorption line strengths (the Lick indices) using stellar
population models, assuming that all of the stars in a galaxy have
identical ages and metallicities. The metallicities and velocity
dispersions have been estimated from spectra of the central regions of
the galaxies.  Since there are radial line-strength gradients in
elliptical galaxies, the central values differ from the global average
values.  Based on the gradients measured by \citet{wu05}, we estimate
that the global total metallicities are typically lower by around 0.25
dex than the central values within 10\% of the effective radius. In
contrast, the $\alpha$/Fe ratio does not appear to show significant
gradients \citep{mehlert03}. Therefore, in comparing observational data
with models, we include downward shifts by 0.25 dex in the observed
values of [$\alpha$/H] and [Fe/H], to correct them to global values, but
assume that the observed central values of [$\alpha$/Fe] are equal to
the global values.

The galaxies in the observational samples are selected in a
heterogeneous way. In order to allow a fairer comparison between
observations and models in Figs.\ref{fig:fig1} and \ref{fig:fig2}, we
have plotted roughly the same number of model galaxies ($\sim$140) as in
the observational sample, randomly selected from a volume-limited model
catalogue so as to match the distribution of velocity dispersion
$\sigma$ found in the observational sample. We note that the model
reproduces the observed $L_B$-$\sigma$ correlation, where $L_B$ is the
total B-band luminosity.

\begin{table}
\begin{center}
\caption{Median of [$\alpha$/H], [Fe/H] and [$\alpha$/Fe] for galaxies
with $150\leq\sigma/{\rm km~s}^{-1}\leq 250$.  Values in the brackets
represent the width between 25\% and 75\% values of the distributions.
For the observations, we show both the measured central values and the
estimated global values. The model values are all global values.  }
\label{tab:tab1}
\begin{tabular}{crrr}
\hline
model & [$\alpha$/H]~~~~~ & [Fe/H]~~~~~ & [$\alpha$/Fe]~~~~~\\
\hline
observations (central) &  0.29 (0.19) &  0.06 (0.21) & 0.23 (0.10) \\
observations (global) &  0.04 (0.19) &  -0.19 (0.21) & 0.23 (0.10)  \\
\hline
 superwind   &  0.04 (0.17) & -0.18 (0.30) & 0.21 (0.14)\\
 superwind/Kennicutt  & -0.16 (0.22) & -0.27 (0.32) & 0.07 (0.10)\\
 conduction & 0.29 (0.16) & 0.09 (0.28) & 0.21 (0.16)\\
\hline
\end{tabular}
\end{center}
\end{table}

Fig.\ref{fig:fig1} shows [$\alpha$/H], [Fe/H] and [$\alpha$/Fe] plotted
against velocity dispersion for the observational sample and for the
three models we consider. The dashed boxes show the parameter region in
which most of the observed elliptical galaxies are found, but in the
panels showing the models, the boxes are plotted taking into account the
estimated correction from central to global metallicities.  To allow a
more quantitative comparison, we also give in Table~\ref{tab:tab1} the
median values and scatter for typical $L_*$ ellipticals, with
$\sigma\approx 200{\rm km~s}^{-1}$.  We note the following points: (1)
The observed ellipticals show a large scatter in both abundances
([$\alpha$/H] and [Fe/H]) and in abundance ratios ([$\alpha$/Fe]) at any
value of $\sigma$, and this scatter is generally reproduced by the
models.  (2) All three models predict similar trends of increasing
[$\alpha$/H] with $\sigma$, which agree with the trend seen in the
observations.  (3) However, the overall normalization of [$\alpha$/H]
differs between the models.  When we take metallicity gradients into
account, the {\em superwind} model (with a top-heavy IMF in bursts)
seems to be in best agreement with the observed [$\alpha$/H] (see
Table~\ref{tab:tab1}). The {\em superwind/Kennicutt} model, which
assumes a solar neighbourhood IMF for all stars, but is otherwise
identical to the {\em superwind} model, predicts the [$\alpha$/H] values
too low by $0.2$ dex. The {\em conduction} model, which also has a
top-heavy IMF in bursts, but has less gas and metals ejected from
galaxies than in the {\em superwind} model, predicts [$\alpha$/H] values
too high by $0.2$ dex.  (4) The models all predict a trend of [Fe/H]
increasing with $\sigma$, in disagreement with the observational data,
which indicate a flat dependence.  (5) The {\em superwind} model
predicts [Fe/H] for $L_*$ galaxies in agreement with observations, when
metallicity gradients are taken into account (see
Table~\ref{tab:tab1}). The {\em superwind/Kennicutt} model predicts only
slightly lower [Fe/H], showing that the top-heavy IMF in bursts affects
the $\alpha$-abundance more than the Fe-abundance. The {\em conduction}
model predicts [Fe/H] too large by nearly 0.3 dex in $L_*$ galaxies.
(6) The trends for [$\alpha$/Fe] vs. $\sigma$ reflect those for
[$\alpha$/H] and [Fe/H] separately. The observations indicate a trend of
[$\alpha$/Fe] increasing with $\sigma$, while the models all predict a
trend in the opposite sense.  Both the {\em superwind} and {\em
conduction} models predict super-solar [$\alpha$/Fe] values for $L_*$
galaxies ($\sigma\approx 200\, {\rm km~s}^{-1}$) which agree well with
observations. However, none of the models match the typical observed
[$\alpha$/Fe] values in high-luminosity ellipticals, with $\sigma\approx
300{\rm km~s^{-1}}$.

The trend we find in our models of the typical [$\alpha$/Fe] decreasing
with increasing velocity dispersion or luminosity is similar to what was
found by \citet{tk99}. However, although \citeauthor{tk99} based their
calculations on star formation histories extracted from hierarchical
models, they computed the chemical evolution treating each galaxy as a
closed box, ignoring galaxy mergers and inflows and outflows of gas and
metals. Our calculation does include all of these effects, so our result
is a stronger one.  We will compare our results with the closed box
result in more detail in a future paper. In all three models presented
here, ellipticals with $\sigma = 200-300 \, {\rm km~s}^{-1}$ had around
40\% of their star formation occuring in bursts, while the most recent
burst typically happened 3-7 Gyr ago.

In Fig.\ref{fig:fig2} we plot the $\alpha$/Fe ratio directly against the
Fe-abundance, for exactly the same samples of observed and model
galaxies as shown in Fig.\ref{fig:fig1}. We see that the observational
data do not show any strong correlation of [$\alpha$/Fe] with [Fe/H],
but instead show a scatter of 0.4 dex in [$\alpha$/Fe] at a given value
of [Fe/H]. The models all show a trend of declining [$\alpha$/Fe] with
increasing [Fe/H], with varying amounts of scatter.  Once we include the
effect of metallicity gradients on the observational data, the {\em
superwind} model seems to be the most consistent with observations,
while the {\em superwind/Kennicutt} and {\em conduction} models predicts
[$\alpha$/Fe] values at a given [Fe/H] which are respectively too low or
too high.

We note here that there is a systematic uncertainty in the
[$\alpha$/H] and [$\alpha$/Fe] values obtained from observational data
due to {\em template bias}. The stellar population models used to
derive metallicities and ages from the observed integrated absorption
line strengths (Lick indices) are based on libraries of observed
stellar spectra, in which some stars have non-solar $\alpha$/Fe
ratios. If no allowance is made for non-solar $\alpha$/Fe in the
template stellar spectra, then this can bias the determination of
[$\alpha$/Fe] and [$\alpha$/H] from the integrated spectra of
galaxies. However, different authors have made different corrections
for this bias, and the issue is currently controversial.
\citet{thomas03} include an internal correction within their
models. \citet{p04} have instead proposed that the following correction should
be applied to [$\alpha$/Fe] values derived from integrated line strengths:
\begin{equation}
 {\rm [\alpha/Fe]_{final}}=-0.5{\rm [Fe/H]+[\alpha/Fe]_{raw}}.
\end{equation}
where ${\rm [\alpha/Fe]_{raw}}$ is the value derived using a stellar
population model which ignores the template bias.  A corresponding
correction is applied to [$\alpha$/H], but no correction is applied to
[Fe/H].  \citet{p04} found that when this correction is applied to
observational data, the [$\alpha$/H]-$\sigma$ relation becomes flatter
but tighter, while the trend of [$\alpha$/Fe] increasing with $\sigma$
almost disappears. However, \citet{tmbd05} have disputed the validity
of this correction and the conclusions of \citeauthor{p04}. Since this
correction is controversial, we have plotted all of the observational
data in Figs.\ref{fig:fig1} and \ref{fig:fig2} (including the data
from \citet{ps02} and \citet{p04}) {\em without} including this
correction. However, to show what the effect of making this correction
would be, we have plotted in both figures dotted boxes indicating
where the observational points would shift to if the correction were
applied. We see that the model predictions in fact agree better with
the corrected than uncorrected data, especially for the {\em
superwind} model.

\section{CONCLUSIONS}
We have investigated the metal enrichment of elliptical galaxies in the
framework of hierarchical galaxy formation models, taking into account
the effects of galaxy mergers and gas inflows and outflows, as well as
the production of metals by both SNe~Ia and SNe~II. This is the first
time this has been done using a semi-analytical model.  Our model is the
same as that which \citet{nlbfc05} used to show that the metallicities
of iron and $\alpha$ elements in the ICM are successfully reproduced
only when a top-heavy IMF is adopted for starbursts. We find that all of
the models we consider predict correlations of the $\alpha$-element
abundance with stellar velocity dispersion $\sigma$ with similar slopes
to what is observed, but the model without a top-heavy IMF in bursts
(the {\em superwind/Kennicutt} model) predicts $\alpha$-element
abundances which are somewhat too low.  Of our two models which include
a top-heavy IMF in bursts, the {\em superwind} model predicts
$\alpha$-element abundances which agree well with observations of bright
elllipticals, while the {\em conduction} model (which has less ejection
of gas and metals) predicts abundances which are too high.

Our models with a top-heavy IMF also predict [$\alpha$/Fe] values in
typical typical $L_*$ ellipticals (with $\sigma \approx 200 \, {\rm
km~s}^{-1}$) which are similar to observed values, while the model in
which all stars form with an IMF like that in the solar neighbourhood
predicts [$\alpha$/Fe] values which are too low. However, none of the
models reproduce the trend of [$\alpha$/Fe] increasing with $\sigma$
which is implied by most observational data.

We note however some caveats to our comparison of the models with
observational data: (i) the observed metallicities are for the central
regions of galaxies, while the models predict global values; (ii)
there are uncertainties in the estimation of metallicities from
stellar absorption line indices connected with the treatment of
non-solar $\alpha$/Fe abundance ratios; (iii) the models used to
estimate metallicities and ages from observed stellar absorption line
indices may give biased results if a galaxy contains stars with a
mixture of ages and metallicities, rather a single stellar population
with a unique age and metallicity. We plan to address these issues in
a future paper, to allow a more detailed test of our models against
observational data on ellipticals.

\section*{ACKNOWLEDGMENTS}    
We acknowledge support from the PPARC rolling grant for
extragalactic astronomy and cosmology at Durham.  MN and TO are
supported by the Japan Society for the Promotion of Science for Young
Scientists (No.207 and 1891).  CMB is supported by a Royal Society
University Research Fellowship. We also thank the referee, Dr. Claudia
Maraston, for a constructive and helpful report.

\bsp
\end{document}